\documentclass[11pt,letterpaper,twoside,reqno]{amsart}
\usepackage{amsmath}
\usepackage{amscd} 
\usepackage[pdftex]{graphicx}
\usepackage{fullpage}
\usepackage{multirow}

\usepackage{url}

\newcommand{\junk}[1]{}
\newcommand{\comment}[1]{}

\newtheorem{theorem}{Theorem}
\newtheorem{definition}[theorem]{Definition}
\newtheorem{criterion}[theorem]{Criterion}
\newtheorem{proposition}[theorem]{Proposition}
\newtheorem{corollary}[theorem]{Corollary}
\newtheorem{lemma}[theorem]{Lemma}

\newcommand{\serr}[1]{\left\| #1 \right\|_{\sim}}
\newcommand{\nerr}[1]{\left\| #1 \right\|}

\newcommand{\diag}{{\rm diag}}
\def \F {\mathbb{F}}
\def \poly {{\rm poly}}

\def \C {\mathbb{C}}

\def \Z {\mathbb{Z}}

\def \K {\mathcal K}

\def \poly {{\rm poly}}

\def \rank {{\rm rank }}

\newcommand\tr{{\rm Tr}}

\newcommand\gr{{\rm gr}}

\def \< {\langle}
\def \> {\rangle}

\newcommand\dotprod[2]{\left\langle {#1},{#2}\right\rangle}
\newcommand\dotprodmag[2]%
  {\left|\left\langle{#1},{#2}\right\rangle\right|}

\newcommand{\halmos}{\hspace*{\fill}\rule{1ex}{1.4ex}}
\makeatletter
\def\newproof#1{\@nprf{#1}}
\def\@nprf#1#2{\expandafter\@ifdefinable\csname #1\endcsname
\global\@namedef{#1}{\@prf{#1}{#2}}\global\@namedef{end#1}{\@endproof}}
\def\@prf#1#2{\@beginproof{#2}{\csname the#1\endcsname}\ignorespaces}
\def\@beginproof#1{\rm \trivlist \item[\hskip \labelsep{\bf #1: }]}
\def\@endproof{\halmos \endtrivlist}
\makeatother

\newtheorem{myalgorithm}[theorem]{Algorithm}
\newenvironment{algorithm}{\begin{myalgorithm}\rm}{%
\nopagebreak\hspace*{\fill}\nopagebreak$\halmos$\end{myalgorithm}}

\title{List decoding of noisy Reed-Muller-like codes}

\author{A.~Robert Calderbank, Anna C.~Gilbert, and  Martin J.~Strauss}

\thanks{A.~Robert Calderbank
  is with the Department of Mathematics, Princeton University,
  Princeton, NJ 08544.  Anna~C.~Gilbert is with the
  Department of Mathematics, The
  University of Michigan at Ann Arbor, 2074 East Hall, 530 Church St.,
  Ann Arbor, MI 48109-1043.  Martin J.~Strauss is jointly appointed
  with the Department of Mathematics and the Department of Electrical
  Engineering and Computer Science at The University of Michigan.
  Calderbank has been supported by DARPA-ONR~N00173-06-1-G006.
  Gilbert and Strauss have been supported by NSF DMS 0354600 and DARPA-ONR~N66001-06-1-2011.}

\email{calderbk@math.princeton.edu, \{annacg,martinjs\}@umich.edu}

\address{Department of Mathematics \\
         Princeton University \\
         Fine Hall \\
         Princeton, NJ 08544}
\address{Department of Mathematics\\
   University of Michigan\\
   2074 East Hall \\
   Ann Arbor, MI 48109}
\address{Depts. of Mathematics and EECS \\
         University of Michigan \\
         Ann Arbor, MI 48109}

\begin{document}

\maketitle

\thispagestyle{empty}

\begin{abstract}
Coding theory has played a central role in the development of computer
science.  One critical point of interaction is decoding
error-correcting codes.  First- and second-order Reed-Muller (RM(1)
and RM(2), respectively) codes are two fundamental error-correcting
codes which arise in communication as well as in
probabilistically-checkable proofs and learning.  In this paper, we
take the first steps
toward extending the quick randomized decoding tools of RM(1) into the
realm of quadratic binary and, equivalently, $\Z_4$ codes.  Our main
algorithmic result is an extension of the RM(1) techniques from
Goldreich-Levin and Kushilevitz-Mansour
algorithms~\cite{GL,kushilevitz91learning} to the {\em Hankel}
code~\cite{CGLMS}, a code between RM(1) and RM(2).  That is, given
signal $s$ of length $N$, we find a list that is a superset of all
Hankel codewords
$\varphi$ with $|\dotprod{s}{\varphi}|^2\ge(1/k)\nerr{s}^2$, in time
$\poly(k,\log(N))$.
We then turn our attention to the widely-studied Kerdock codes.  We
give a new and simple formulation of a known Kerdock code as a
subcode of the Hankel code.  We then get two immediate corollaries.
First, our new Hankel list-decoding algorithm covers subcodes,
including the new Kerdock construction, so we can list-decode Kerdock,
too.  Furthermore, exploiting
the fact that dot products of distinct Kerdock vectors have small
magnitude, we get a quick algorithm for finding a sparse Kerdock
approximation.  That is, for $k$ small compared with $1/\sqrt{N}$ and
for $\epsilon>0$, we
find, in time $\poly(k\log(N)/\epsilon)$, a $k$-Kerdock-term
approximation
$\widetilde s$ to $s$ with Euclidean error at most the factor
$(1+\epsilon+O(k^2/\sqrt{N}))$ times that of the best such
approximation.
\end{abstract}

\section{Introduction}

Coding theory and computation have enjoyed a long and fruitful
interaction.  Decoding a received codeword is inherently an
algorithmic problem and, conversely, codes have been used as key
components of algorithms for many purposes, including
pseudorandomness, probabilistically checkable proofs, learning, and
cryptography.  The computational view of codes can also provide
important insights for coding theory and code construction.
See~\cite{madhu-hp,sudan:survey} and the references therein for a
sample of this fruitful interaction.

Because decoding is inherently an algorithmic problem, it is natural
to analyze the computational cost of decoding a received codeword.  We
can quantify how much time and space we need to decode a vector which
has been corrupted according to a variety of noise models.  In this
paper, we are interested in how many samples of the received codeword
are necessary for decoding, how much noise we can tolerate in the
input, and how quickly we can decode using just a few random samples
in the presence of this noise.

The first- and second-order binary Reed-Muller codes RM(1) and RM(2)
are fundamental in the study of codes and their applications to
algorithms.  A RM(1) codeword of dimension $n$ can be regarded as a
binary linear function on $n$ variables and a RM(2) codeword is a
quadratic function on $n$ variables.  As such, they are fundamental
expressive classes, used in proofs and learning as well as error-free
communication.

Binary RM(1), in particular, admits highly efficient algorithms for
decoding, even in the presence of noise.  We are interested in a form
of decoding that has appeared many times before with various names,
and that we call {\em Euclidean List Decoding}.
The first quick algorithms for Euclidean list decoding of RM(1) are
in~\cite{GL,kushilevitz91learning}.  Given a 
(multiplicatively-written) linear function $f:\Z_2^n\to(\pm 1,\cdot)$,
one can recover $f$ by querying its value on just $\poly(n)$ values of
its graph, instead of all $2^n$ values.  Furthermore, the decoding
succeeds even in the presence of a lot of noise; {\em i.e.}, if the
noise $\nu$ is orthogonal to the signal $f$ and if we assume only that
$\nerr{f}^2\ge(1/k)\nerr{\nu}^2$, then the algorithm on $f+\nu$ takes
time polynomial in $kn$ and returns a (short) list of possible
$f$'s.  See~\cite{sudan:list} for
a discussion of list decoding algorithms and their applications.  We
note that this problem can be solved more generally using
nearest-neighbor data
structures~\cite{Ind00:High-Dimensional-Computational}, but the
general solution requires space and preprocessing time $N=2^n$, which we
want to avoid.

While the available techniques for RM(1) make it useful in many
applications, RM(1) is limited in several important ways compared with
RM(2).  First, there
are only $2^n$ RM(1) codewords, while there are approximately
$2^{n^2/2}$ RM(2) codewords, so, quantitatively, RM(2) is
more expressive.  But there are important structural differences, as
well.  When used to express a concept or to code a computation, an
RM(1) codeword as a function considers its variables one at a time, while RM(2)
codewords consider their variables in pairs.  First-order Reed-Muller
codewords form an orthonormal basis, while RM(2) forms a highly
redundant {\em dictionary}---a collection of more than $N$ vectors
spanning a vector space of dimension $N$---that is potentially much
more useful for
lossy compression.  When used as a pseudorandom number generator, RM(1)
provides a family of 3-wise independent random variables  and RM(2)
provides a family of 7-wise random variables.\footnote{That
is, if we fix any three indices $y_1,y_2,y_3$ into an unknown codeword
$\varphi$ and then choose an RM(1) codeword $\varphi$ at random, the
random variables $\varphi(y_1),\varphi(y_2),\varphi(y_3)$ are jointly
independent.  If we choose an RM(2) codeword at random, any 7
positions are independent.}  Because of the extra
expressiveness of RM(2), however, many tools from the first order
theory do not apply.  For example, we do not know how to recover a
RM(2) vector in the presence of noise unless the noise is
slight~\cite{alon:kaufman}.

In this paper, we take the first steps toward extending the decoding
tools of RM(1) into the realm of quadratic binary (and, equivalently,
$\Z_4$) codes.  We show how to recover {\em Hankel} codewords~\cite{CGLMS}
efficiently in the
presence of noise, giving a result analogous to what one can do with
RM(1) up to a polynomial in the parameters.  The Hankel code is the
union of {\em cosets} of RM(1), {\em i.e.}, $\bigcup_{\varphi\in
  Q}\varphi\mbox{RM(1)}$ for some $Q$ of size $q$, so that Hankel can
be regarded as the
union of $q$ orthonormal bases, each equivalent to RM(1).  It follows
immediately that one can use the
\cite{kushilevitz91learning}~algorithm $q$ times to do
list-decoding over the union of $q$ equivalent copies of RM(1), but
only at time cost $q$ times the cost of one instance of the algorithm
in~\cite{kushilevitz91learning}.  Hankel
consists of $q=\Theta(N^2)$ copies of RM(1), however, so the cost of
such a trivial algorithm would be prohibitive.  By contrast, we
list-decode Hankel
in total time $\poly(k,\log(N))$.  Such efficient list-decoding is
possible only by confluence of the choice of dictionary (Hankel) and
the algorithm, and represents an important way in which our
contribution is significant.

We also give a new, simple construction of a code in the
well-studied class of {\em Kerdock} codes.  Our Kerdock construction
$\mathcal{K}$ is a subcode of the Hankel code $\mathcal{H}$, which
implies immediately that our Hankel list-decoding algorithm applies
also to our Kerdock construction.
Thus we have
RM(1)$\subseteq\mathcal{K}\subseteq\mathcal{H}\subseteq$RM(2).  While
Kerdock and Hankel are still in some important respects more limited
than RM(2), they are great improvements over RM(1).  For example, a
random codeword from a Kerdock code (and, therefore, from the Hankel
code) provides a family of 5-wise independent random variables.  Each
Kerdock code has $N^2$ vectors and the Hankel code has $\Theta(N^3)$
vectors, compared with $\Theta(N)$ for RM(1) and
$2^{\Theta(\log^2(N))}$ for RM(2).  Kerdock
represents a substantial, well-studied family of quadratic functions
with advantages over RM(1) in the areas of coding
theory~\cite{hammons:calderbank}, radar
signaling~\cite{howard:calderbank:moran}, and spread-spectrum
communication.

Finally, the previous work
in~\cite{TGMS03:Improved-Sparse,gilbert03approximation}
demonstrates that we can use a fast list-decoding algorithm for
to find 
a {\em sparse representation} efficiently.  That is, exploiting
the fact that dot products of distinct Kerdock vectors have small
magnitude, we get a quick algorithm for finding a sparse Kerdock
approximation.  More specifically, for any $k<1/(6\sqrt{N})$ and any
$\epsilon>0$, we can find,
in time $\poly(k,\log(N),1/\epsilon)$, a $k$-Kerdock-term
approximation $\widetilde s$ to $s$ with Euclidean error at most the
factor $(1+\epsilon+O(k^2/\sqrt{N}))$ times that of the best such
approximation.

This paper is organized as follows.  In Section~\ref{sec:prelim}, we
give preliminaries about finite fields, Reed-Muller codes, and Kerdock
codes.  We also include a discussion of related work.  In
Section~\ref{sec:newK}, we give a new, computational construction of a
Kerdock code, as a subcode of the Hankel code.  In
Section~\ref{sec:recovery}, we give our algorithm for fast list
decoding of the Hankel code.  In Section~\ref{sec:conc}, we give
corollaries of our main result concerning list-decoding and sparse
recovery of Kerdock codes, as well as indications about directions for
improvement.

\section{Preliminaries}\label{sec:prelim}
\subsection{Finite fields}
To outline the setting in which Kerdock codes are defined, we begin
with the definition of finite fields and the algebra we perform over
these fields.  Let $h(t)$ be a polynomial of degree $n$ over
$\Z_2$ that is primitive, {\em i.e.}, $h(t)$ does not divide $t^k - 1$
for any $k < 2^n -1$.  Because $h$ is a primitive (and hence, irreducible)
polynomial, it has no non-trivial factorization.

The ring of polynomials $\Z_2[t]$ modulo $h$, $\Z_2[t]/h$, forms a
field of $2^n$ elements.  We denote this field $\F(2^n)$.  The
polynomial $\xi(t) = t$ is a (multiplicative) generator of the field;
thus, the set $\{1, \xi, \xi^2, \ldots, \xi^{2^n-1}\}$
enumerates the non-zero elements of the field.  Additively, the field
$\F(2^n)$ is a vector space $\Z_2^n$ over $\Z_2$ of dimension $n$ with basis
$\{1, \xi, \xi^2, \ldots, \xi^{n-1}\}$.  It is also a quotient vector space
of $\Z^n$.  When we want to emphasize the
vector formulation of a field element $\alpha$, we write
$[\alpha]$ for a column vector.  Thus
$[1],[\xi],[\xi^2],\ldots,[\xi^{n-1}]$ are the canonical basis
vectors.  Below, we will often want to consider these
$\{0,1\}$-valued vectors to be in $\Z_2^n,\Z_4^n$, or $\Z^n$ for the
purposes of dot products.  We will write, {\em e.g.}, $i^{[y]^TQ[y] +
  2\ell^T[y]}$, where $y$ is a field element, $Q$ is a $\{0,1\}$-valued matrix,
and $\ell$ is a $\{0,1\}$-valued vector.  Note that all the arithmetic in the
exponent can be done over $\Z$, where $[y]$ is a $\{0,1\}$-valued vector.
Since the exponent is an exponent of $i$, arithmetic can equivalently
be done mod 4.  Finally, since 2 multiplies $\ell^T[y]$, the dot product
of $\ell$ and $[y]$ can be performed mod 2.
For any $x \in \F(2^n)$, we have $x^{2^n} = x$ so that
$\sqrt x = x^{2^{n-1}}$.  Because 2 is congruent to 0 mod 2, we have
$(x + y)^2 = x^2 + y^2$ for any $x, y \in \F(2^n)$ and, by repeated
squaring, $(x + y)^{2^j} = x^{2^j} + y^{2^j}$.

The trace of an element $x \in \F(2^n)$ is an important quantity we
use in defining and constructing Kerdock codes.
\begin{definition}
The trace of $x \in \F(2^n)$, $\tr(x)$, is defined to be
\[ \tr(x) = \sum_{0 \leq j < n} x^{2^j} = x + x^2 + \cdots +
            x^{2^{n-1}}.
\]
\end{definition}
The following lemma gives the properties we need of the trace map.  We
give the simple proof for completeness.
\begin{lemma}
We have
\begin{itemize}
\item For $x,y\in\F(2^n)$ and
  $a,b\in\F(2)$, we have $\tr(ax+by)=a\tr(x)+b\tr(y)$.
\item The image of $\tr$ is in $\Z_2$.
\item The trace is not identically 0.
\end{itemize}
\end{lemma}
\begin{proof}
(Repeated) squaring of an element is a linear operator, so $\tr(ax
+ by) = a \tr(x) + b \tr(y)$.

Again by linearity of squaring, $\tr(x^2)=\tr(x)^2$.  Since
$x^{2^n}=x$, we have $\tr(x)=\tr(x^2)=\tr(x)^2$.  Thus $\tr(x)$
satisfies $y=y^2$, whence $\tr(x)\in\F(2)$.  For $n$ odd, $\tr(1)=1$.
(Lemma~\ref{lem:fullrank} shows that $\tr\not\equiv 0$ for even $n$,
as well.)
\end{proof}

Thus $\tr$ is an additive homomorphism from the big field $\F(2^n)$ to
the {\em prime subfield} $\F(2)=\Z_2$, so $\tr(x)=0$ for exactly half
of the field elements.  It is not necessarily true that
$\tr(xy)=\tr(x)\tr(y)$.  Finally, note that $\tr(1)$ is 0 or 1 if $n$
is even or odd, respectively.

\subsection{Definitions of RM(1,$n$) and RM(2,$n$)}
We review the definitions of the two codes, first- and second-order
Reed-Muller codes (RM(1,$n$) and RM(2,$n$), respectively), which
sandwich Kerdock codes.  Fix a parameter $n$.
\begin{definition}
Let $\ell \in \Z_2^n$ be a binary vector of length $n$ and let $\epsilon
\in \Z_2$.  The first-order Reed-Muller code RM(1,$n$) of length $N =
2^n$ is defined as a set of vectors $v_{\ell,\epsilon}$ indexed by $\ell$
and $\epsilon$.  For each code word $v_{\ell, \epsilon}$ at position $[y]
\in \Z_2^n$ is given by
\[    v_{\ell,\epsilon}([y]) = 2 ( \ell^T [y] + \epsilon) \mod 4.
\]
\end{definition}
The exponentiated form of RM(1,$n$) is given by
\[   \varphi_{\ell,\epsilon}([y]) = \frac{1}{\sqrt N} i^{2(\ell^T [y] + \epsilon)} = 
     \frac{(-1)^{\ell^T[y] + \epsilon}}{\sqrt N}.
\]
We normalize the codevectors by $\sqrt N$ in the exponentiated form to
obtain unit vectors.

\begin{definition}
Let $Q$ be an $n \times n$ symmetric matrix over $\Z_2$, let $\ell \in
\Z_2^n$ be a binary vector of length $n$, and let $\epsilon \in
\Z_4$.  The second-order Reed-Muller code RM(2, $n$) of length $N =
2^n$ is defined as a set of vectors $w_{Q,\ell,\epsilon}$ indexed by $Q$,
$\ell$, and $\epsilon$.  Each codeword $w_{Q,\ell,\epsilon}$ at position $[y]
\in \Z_2^n$ is given by
\[   w_{Q,\ell,\epsilon}([y]) = ([y]^T Q [y] + 2 \ell^T [y] + \epsilon) \mod 4.
\]
\end{definition}
The exponentiated form of RM(2,$n$) is given by
\[  \varphi_{Q,\ell,\epsilon}([y]) = \frac{1}{\sqrt N}
                i^{[y]^T Q [y] + 2 \ell^T [y] + \epsilon}.
\]
Below, we will sometimes abbreviate the index $(Q,\ell,\epsilon)$ as
$\lambda$, so that $\varphi_{Q,\ell,\epsilon}=\varphi_\lambda$.
Again, we normalize the codevectors in the exponentiated form so they
are unit vectors.  Observe that if $Q = 0$, then the subset of
RM(2,$n$) codewords given by $w_{0,\ell,\epsilon}$ are, in fact,
RM(1,$n$) codewords.  We frequently drop the index $\epsilon$
since $i^\epsilon$ represents a unit factor that can be absorbed into
a more general coefficient $c_{Q,\ell}$ of
$c_{Q,\ell}\varphi_{Q,\ell}$.

In other literature, both RM(1) and RM(2) are presented
as binary codes.  Our theory can be formulated for both $\Z_2$ and
$\Z_4$, but we stick to $\Z_4$ after giving the equivalence between
previous work and ours.  We will
consider RM(1) and RM(2) over $\Z_4$, as above, since the Kerdock codes
are most natural over $\Z_4$---they are nonlinear binary codes but
linear over $\Z_4$.

We say
that a code with entries in $\Z_4$ is a $\Z_4$-code while one with
entries in $\Z_2$ is a $\Z_2$-code.  The two previous definitions of
RM(1,$n$) and RM(2,$n$) both result in $\Z_4$-codes.  The $\Z_2$
Reed-Muller codes may be more familiar to the reader and we often want
to relate a $\Z_4$-code to a $\Z_2$-code. We do so via the Gray map.
\begin{definition}
The Gray map, $\gr: \Z_4 \to \Z_2^2$, is given by
\[    \gr(0) = 00, \quad \gr(1) = 01, \quad \gr(2) = 11 \quad\text{and}\quad
      \gr(3) = 10.
\]
\end{definition}
We sometimes use the exponential version, from $\{\pm1,\pm i\}$ to
$(\pm1)^2$, given by
\[   \gr(+1)=(+1,+1), \quad \gr(+i)=(+1,-1),\quad \gr(-1)=(-1,-1),
\quad\text{and}\quad \gr(-i)=(-1,+1).
\]
Further overloading notation, $\gr:Z_4^N\to\Z_2^{N\times 2}$ is gotten
by applying the Gray map to each of $N$ elements in a vector in
$\Z_4^N$, getting $N$ elements in $\Z_2^2$, and similarly for the
exponential versions.

Equivalently, one can transform $Q$, a $\Z_4$-valued
quadratic form on $\Z_2^n$, to $M$, a $\Z_2$-valued quadratic form on
$\Z_2^{n+1}$.  The quadratic form $Q$ is an $n \times n$ binary
symmetric matrix while $M$ is an $(n+1) \times (n+1)$ binary skew
symmetric matrix---that is, $M$ has zero diagonal.  Let the row vector
$d_Q$ be the diagonal of
$Q$.  Then Calderbank et al.~\cite{calderbank:cameron:kantor:seidel}
show that the correspondence between binary symmetric matrices $Q$ and
binary skew symmetric matrices $M$ is given by 
\begin{equation}\label{eq:z2z4}
    M = \begin{pmatrix}
          0 & d_Q^T \\
          d_Q & d_Q d_Q^T + Q \\
      \end{pmatrix},
\end{equation}
where the ``extra'' bit in the top row and left column is used as an
index into the two outputs of the Gray map.
This correspondence is not linear but it is rank preserving in the
sense that if $M$ has rank $n + 1 - 2j$ then $Q$ has rank $n + 1 - 2j$
or $n - 2j$ for any integer $j$, $0 \leq j < (n-1)/2$.

In summary, the following commutative diagram relates codewords and
codeword labels in the $\Z_2$ and $\Z_4$ formulations:
\[
\begin{CD}
\text{$\Z_4$ label}     @>\text{(\ref{eq:z2z4})}>> \text{$\Z_2$ label}\\
@VVV                                 @VVV\\
\text{$\Z_4$ codeword}  @>\text{Gray map}>>        \text{$\Z_2$ codeword}
\end{CD}
\]

The following theorem of Calderbank et
al.~\cite{calderbank:cameron:kantor:seidel} relates the rank of the
binary symmetric matrices $Q_1$ and $Q_2$ to the magnitude of the dot
product between two codewords generated with the respective matrices.
\begin{theorem}\label{thm:dickson}
Let $Q_1$ and $Q_2$ be binary symmetric $n \times n$ matrices and let
$\varphi_{Q_1,\ell_1,\epsilon_1}$ and
$\varphi_{Q_2,\ell_2,\epsilon_2}$ be distinct
exponentiated $\Z_4$-RM(2,$n$) codewords.  If Rank$(Q_1 - Q_2) = R$,
then
\[   \dotprodmag{\strut \varphi_{Q_1,\ell_1,\epsilon_1}}
      {\varphi_{Q_2,\ell_2,\epsilon_2}} \in 
      \left\{\strut 0, 2^{-R/2}\right\}.
\]
In particular, if $\ell_1 = \ell_2$, then the magnitude of the dot product
is $2^{-R/2}$.
\label{thm:dotprod}
\end{theorem}

\subsection{Definition of Kerdock codes}
A Kerdock {\em code} is associated with a Kerdock {\em set} of
matrices.  The definition of the latter is non-constructive.
\begin{definition}
A {\em Kerdock set} $\K$ is a set of $n \times n$ binary symmetric matrices,
including zero, of size $n$ such that for any distinct $P_1, P_2 \in \K$,
the rank of $(P_1 + P_2)$ over $\F(2)$ is $n$.
\end{definition}

In particular,
any non-zero $P \in \K$ has full rank.  We take these matrices $P$ to
be quadratic forms over $\Z_4$.
Each Kerdock set has size at most $N=2^n$, since distinct elements of
a Kerdock set must have distinct top rows.  In fact, Kerdock sets can
achieve maximal size (see below).

\begin{definition}
A Kerdock code $K(n)$ of length $N = 2^n$ is defined as a set of
vectors $c_{P,\ell,\epsilon}$, indexed by $P$, $\ell$, and $\epsilon$.  Each
codeword $c_{P,\ell,\epsilon}$ at position $[y] \in \Z_2^n$ is given by
\[       c_{P,\ell,\epsilon}([y]) = ([y]^T P [y] + \ell^T [y] + \epsilon) \mod 4
\]
where $P \in \K$ comes from a Kerdock set.
\end{definition}

\subsection{Related Work}

The work most closely related to our decoding algorithm is that
of~\cite{GL,kushilevitz91learning},
which was already discussed.  Similar sparse decoding of the Fourier
basis (over $\Z_N$, not $\Z_2^n$) was given
in~\cite{Mansour,stoc-Fourier, akavia:goldwasser}.  Other work on
local testing of codes~\cite{Kaufman05} focuses on limiting the number
of samples, but not the runtime.  Still other work on list-decoding of
Reed-Muller codes~\cite{sudan:survey} focuses on large alphabets,
whereas we work over $\Z_2$.  The problem of {\em testing} low-degree
polynomials~\cite{arora97improved} is different from {\em decoding},
which is what we do for special quadratic polynomials.  We note
that~\cite{alon:kaufman}, in addition to giving lower bounds on the
number of samples for testing binary Reed-Muller codes, also give a
decoding algorithm for a single Reed-Muller vector in the presence of
very small noise.

As for construction of Kerdock codes, the history is as follows.
Kerdock codes were first defined~\cite{macwilliams:sloane}
non-constructively in terms of the allowable quadratic forms.
Later, in the breakthrough
paper~\cite{calderbank:cameron:kantor:seidel}, the authors give
algebraic
constructions of Kerdock codes that provide a rich set of
symmetries, but the algebra included theory somewhat beyond finite
fields.  Independent of and somewhat earlier than the publication of
our work, a construction of a Kerdock code similar to ours is given
in~\cite{HSP}.  Both the construction in~\cite{HSP} and our
construction here are isomorphic, in some sense, to a
construction in~\cite{calderbank:cameron:kantor:seidel}.
We believe our construction is a bit simpler than
~\cite{HSP}---indeed, to get the
Hankel structure we need here, it is simpler for us to give
Definition~\ref{def:lf-k} (below) from scratch than to adapt the construction
in~\cite{HSP}.  As additional value
beyond~\cite{HSP}, we also contribute a
self-contained proof of correctness of the construction, simplifying
the proof in~\cite{calderbank:cameron:kantor:seidel} (\cite{HSP}
gives offers no new proof of correctness).  We also give an important
new characterization of the construction, Lemma~\ref{lem:commute}.

\section{New definition of Kerdock codes}\label{sec:newK}
In this section, we present a construction of Kerdock codes.  

\subsection{Kerdock matrices}\label{sec:kmatrices}
In each construct, a Kerdock code of length $N = 2^n$ is a subset of
RM(2,$n$) code $\{\varphi_{Q,\ell}\}$ satisfying an appropriate
restriction on the binary symmetric
matrix $Q$.  We call these matrices Kerdock matrices.  Roughly
speaking, they are a restricted set of binary Hankel\footnote{A Hankel
  matrix is constant along reverse diagonals.} matrices where the top
row of the Hankel matrix consists of arbitrary entries and each of the
remaining reverse diagonals is gotten from a fixed linear combination
of the previous $n$ reverse diagonals.

Let $h(t) = h_0 + h_1 t + \cdots h_{n-1} t^{n-1} + t^n$ be a primitive
polynomial over $\Z_2$ of degree $n$.  The coefficients of this
polynomial are the coefficients in our fixed
linear mapping.
\begin{definition}\label{def:lf-k}
An $n \times n$ linear-feedback-Kerdock matrix (briefly, lf-Kerdock
matrix) is a Hankel matrix where the top row of
the matrix $a_0, a_1, \ldots, a_{n-1}$ consists of $n$ arbitrary
values in $\Z_2$ and the $j$th reverse diagonal parameter for $j \geq n$ is a
fixed linear combination of the previous $n$ reverse diagonal parameters, given
by
\[   a_j = \sum_{0 \leq \ell < n} a_{j - n + \ell} h_{\ell}.
\]
\end{definition}
(See Section~\ref{sec:example} for an example.)  We denote by $\K$ the
set of lf-Kerdock
matrices.
Next, we provide what turns out to be an equivalent definition of
lf-Kerdock matrices, called trace-Kerdock matrices.
\begin{definition}
An $n$-by-$n$ {\em trace-Kerdock matrix} $K_\alpha$ is the matrix
whose $(j,k)$ position is $\tr(\alpha\xi^{j+k})$ for some $\alpha$
in $\F(2^n)$.
\end{definition}

Note that the set of trace-Kerdock {\em matrices} is $\Z_2$-linear,
meaning the sum (mod 2) of two trace-Kerdock matrices is itself a
trace-Kerdock matrix, by additivity of the trace.  Kerdock {\em
codes}, however, are $\Z_4$-linear but {\em not} $\Z_2$-linear.  

\begin{lemma}\label{lem:fullrank}
Trace-Kerdock matrices have full rank.
\end{lemma}
\begin{proof}
Given trace-Kerdock matrix $K_\alpha$ for $\alpha\ne 0$, regard it as
a matrix over $\F(2^n)$.  Since $K_\alpha$ consists of $0$'s and
$1$'s, the determinant of $K_\alpha$ over $\F(2^n)$ is the same as
the determinant over $\Z_2$.

The matrix $K_\alpha$ factors as $K_\alpha=V^TD_\alpha V$, where
\[
V=
\left(
\begin{array}{cccccc}
1 & \xi   & \xi^2 & \xi^3    & \cdots & \xi^{n-1}\\
1 & \xi^2 & \xi^4 & \xi^6    & \cdots & \xi^{2(n-1)}\\
1 & \xi^4 & \xi^8 & \xi^{12} & \cdots & \xi^{4(n-1)}\\
\vdots &  &       &          &        &  \\
1 & \xi^{2^k} & \xi^{2^k\cdot 2} & \xi^{2^k\cdot 3} & \cdots & \xi^{2^k\cdot(n-1)}\\
\vdots &  &       &          &        &  \\
1 & \xi^{2^{n-1}} & \xi^{2^{n-1}\cdot 2} & \xi^{2^{n-1}\cdot 3} &
\cdots & \xi^{2^{n-1}\cdot(n-1)}
\end{array}
\right)
\]
is vandermonde and
$D_\alpha=\diag(\alpha,\alpha^2,\alpha^4,\alpha^8,\ldots,\alpha^{2^{n-1}})$.
Over the big field,
\[\det(D_\alpha)=\alpha^{1+2+4+\cdots+2^{n-1}}=\alpha^{2^n-1}=1\]
and
the vandermonde parameters $\xi,\xi^2,\xi^4,\ldots$ are distinct, so
$V$ is non-singular.  It follows that
$\det(K_\alpha)\ne 0$ over the field, so $\det(K_\alpha)=1$.
\end{proof}

Note that the factorization $K_\alpha=V^TD_\alpha V$ also shows that
$K_x J K_y J =
K_{xy} J$, where $J=(V^TV)^{-1}=K_1^{-1}$.  Thus the map $x\mapsto K_x
J$ is a non-trivial multiplicative homomorphism from field elements to
matrices.  Since squaring is linear, $x\mapsto D_x$ and, so,
$x\mapsto K_x J = V^T D_x (V^T)^{-1}$ are additive homomorphisms.  It
follows that $x\mapsto K_x J$ is a field homomorphism, so, in
conjunction with the matrix $J$, the trace-Kerdock matrices can be
regarded as field elements.

\begin{lemma}
Every trace-Kerdock matrix is a lf-Kerdock matrix.
\end{lemma}

\begin{proof}
Fix a trace-Kerdock matrix $K_\alpha$.  It is Hankel by
inspection, since the matrix entry $K_\alpha(j,k)=\tr(\alpha\xi^{j+k})$ depends only on
$j+k$.  Note that the
first $n$ diagonal parameters are given by
\[\tr(\alpha),\tr(\alpha\xi),\tr(\alpha\xi^2),\ldots,\tr(\alpha\xi^{n-1}).\]

Fix $j$ and $k$ with $j<n, k<n$, and $j+k\ge n$.  Then the $j+k$
reverse diagonal of
$K_\alpha$, which can be taken mod 2, is
$[\xi^j]^TK_\alpha[\xi^k]=\tr(\alpha\xi^{j+k})$.  Using
additivity of the trace,
\begin{eqnarray*}
\tr(\alpha\xi^{j+k})
& = & \tr(\alpha\xi^{j+k-n}\xi^n)\\
& = & \tr\left(\alpha\xi^{j+k-n}\sum_{\ell<n}h_\ell\xi^\ell\right)\\
& = & \sum_{\ell<n}h_\ell\tr\left(\alpha\xi^{j+k-n+\ell}\right).
\end{eqnarray*}
That is, the $(j+k)$'th reverse diagonal depends linearly on the
previous $n$, for $j+k\ge n$.
\end{proof}

\begin{lemma}
Every lf-Kerdock matrix is trace-Kerdock.
\end{lemma}
\begin{proof}
There are $2^n$ lf-Kerdock matrices since the top row of $n$ bits
enumerates $\Z_2^n$.  There are $2^n$ trace-Kerdock matrices
$K_\alpha$ since the top-left entry in $D_\alpha=(V^T)^{-1}K_\alpha
V^{-1}$ enumerates $\F(2^n)$.  So there
are equal numbers of lf-Kerdock and trace-Kerdock matrices.  Above we
showed that every trace-Kerdock is a lf-Kerdock.  Our statement follows.
\end{proof}

Thus we have
\begin{theorem}
The set of lf-Kerdock matrices is a maximal lf-Kerdock set.
\end{theorem}

Henceforth, we refer to lf-Kerdock and trace-Kerdock matrices as
``Kerdock matrices.''  As above, a Kerdock code is defined from a
Kerdock set $\mathcal K$ as $\{\varphi_{P,\ell}:P\in\mathcal K,
\ell\in\Z_2^n\}$.

\subsection{Example Kerdock matrix construction}\label{sec:example}

Let $n = 3$.  The polynomial $h(t) = 1 + t^2 + t^3 = h_0 + h_2 t^2 +
t^3$ is a primitive polynomial over $\Z_2$ of degree 3.  A $3 \times
3$ Kerdock matrix
\[    P = \begin{pmatrix}
          a_0 & a_1 & a_2 \\
          a_1 & a_2 & a_3 \\       
          a_2 & a_3 & a_4 \\
          \end{pmatrix}
\]
has five reverse diagonal parameters, $a_0,\ldots,a_4$.
We construct $P \in \K$ by choosing the top row $\begin{pmatrix}a_0 &
  a_1 & a_2\\ \end{pmatrix}$ arbitrarily, {\em e.g.}, $\begin{pmatrix}a_0 &
  a_1 & a_2\\ \end{pmatrix} = \begin{pmatrix}1 & 1 &
  1\\ \end{pmatrix}$.  The two remaining reverse diagonals
$a_3$ and $a_4$ are given by
\[ 
    a_3 = a_0 + a_2 = 1 + 1 = 0
    \quad\text{and}\quad
    a_4 = a_1 + a_3 = 1 + 0 = 1.
\]
This results in the matrix
\[    P = \begin{pmatrix}
          1 & 1 & 1 \\
          1 & 1 & 0 \\       
          1 & 0 & 1 \\
          \end{pmatrix}.
\]

\subsection{Properties of Kerdock codes}

We now give a lemma that will be useful in Section~\ref{sec:conc}
as well as in its own right.

\begin{lemma}\label{lem:commute}
Fix a primitive polynomial $h$ for defining a finite field and for the
Kerdock properties.  Let $P$ be a symmetric matrix.  The following are
equivalent:
\begin{itemize}
\item $P$ is Kerdock;
\item For all $r$ and $s$, we have
  $[r]^T P [s] =[\sqrt{rs}]^T P [\sqrt{rs}]$ mod 2;
\item For all $x,y$ and $z$ we have $[x]^T P [yz] = [xy]^T P [z]$ mod 2.
\end{itemize}
\end{lemma}
\begin{proof}
First we show that the two algebraic statements are equivalent.
Suppose $[x]^T P [yz] = [xy]^T P [z]$ holds for all $x,y$, and $z$.
Then, given non-zero $r$ and $s$, put $x=r,y=\sqrt{s/r}$, and
$z=\sqrt{rs}$; it follows that $[r]^T P [s] = [\sqrt{rs}]^T P
[\sqrt{rs}]$.  Conversely, if $[r]^T P [s] =[\sqrt{rs}]^T P
[\sqrt{rs}]$ for all $r$ and $s$, then, given $x,y,z$, we have $[x]^T
P [yz] = [\sqrt{xyz}]^T P [\sqrt{xyz}] = [xy]^T P [z]$, first 
putting $r=x$ and $s=yz$ and then putting $r=xy$ and $s=z$.

Now, suppose $P$ is Kerdock and fix $x,y$, and $z$.  By linearity, it
suffices to consider $x=\xi^j$ and $z=\xi^k$, for $0\le j,k<n$.
Because $\xi$ is a multiplicative generator, it suffices to consider
$y=\xi$.  If $j<n-1$ and $k<n-1$, then $[x]^T P [yz] = [xy]^T P [z]$ follows
from Hankelness.  If $j<n-1$ and $k=n-1$, then
$[xy]^T P [z] =[\xi^{j+1}]^T P [\xi^{n-1}]$.  By the linear feedback
Kerdock property, this
equals $\sum_{\ell<n}h_\ell[\xi^j]^T P [\xi^\ell]$.  By linearity, this is
$[\xi^j]^T P [\sum_{\ell<n}h_\ell\xi^\ell]$.  By definition of $h$,
this is $[\xi^j]^T P [\xi^n] = [x]^T P [yz]$, as desired.  A similar
analysis holds if $j=n-1$ and $k<n-1$.  The case $j=k=n-1$ follows
from symmetry of $P$.  

Conversely, suppose $[x]^T P [yz] = [xy]^T P [z]$ for all $x,y$, and
$z$.  Consider the $j$'th row of $P$, for $j>0$.  We want to show that
it is gotten by shifting the $j-1$'st row to the left and setting the
rightmost entry of the $j$'th row to the appropriate linear
combination of the items in the $j-1$'st row.  Put $x=\xi^j,y=\xi$,
and $z=\xi^k$.  Then, for $k<n-1$, Hankelness (and, therefore, the
statement) follows immediately.  For $k=n-1$, we have
$yz=\xi^n=\sum_{\ell<n}h_\ell\xi^\ell$, and the statement follows by
additivity of the
trace.
\end{proof}

\subsection{Kerdock and random variables of limited independence}

We first give a definition of limited independence for a family of
random variables.  This is satisfied by the positions in a random
Kerdock codeword.

\begin{definition}
A $\Z_4$-code is 3.5-wise independent if the distribution on any three
positions of a random codeword is uniform on $(\Z_4)^3$ and,
conditioned on any three positions, for any fourth position $X$, we
have $\Pr(X=0)=\Pr(X=2)$ and $\Pr(X=1)=\Pr(X=3)$.
\end{definition}

For example, in our construction, the joint
distribution is uniformly random conditioned on the sum of the four
positions being 0 or 2 mod 4.

The notion of 3.5-wise independence is useful because it can
substitute for 4-wise independence in some cases, even when 3-wise
independence cannot.  Consider the exponentiated version of of a
3.5-wise independent family, so each value is $\pm1$ or $\pm i$.  Then
any four random variables $W,X,Y,Z$ satisfy that $W,X,Y$ are
independent and, conditioned on $W,X,Y$, the expectation of $Z$ is 0,
because $Z=\pm1$ uniformly conditioned on $Z$ real and $Z=\pm i$
uniformly conditioned on $Z$ imaginary; the probability that $Z$ is
real is arbitrary.
Thus, in a family of $N$ 3.5-wise independent
random variables, the first four moments agree (up to constant
factors) with the moments of a truly random family.  We have
$E[|\sum X_j|^k]=\Theta(N^{k/2})$ for $k = 
0,2,4$ and $E[(\sum X_j)^k]=0$ for $k=1,3$.  For the 3-wise
independent family RM(1), we have $E[|\sum X_j|^4]=\Theta(N^3)$,
since, for
any triple $W,X,Y$ of variables, there is exactly one fourth variable
$Z$ such that $WXYZ\equiv 1$ and all other 4-tuples have zero
expectation.

The following had been known, but not previously presented in terms of
3.5-wise independence.
\begin{lemma}
For odd $n\ge 3$, the Kerdock code is 3.5-wise independent but not
4-wise independent.  For even $n\ge 3$, the Kerdock code is 3-wise but
not 3.5-wise independent.
\end{lemma}

\begin{lemma}
For any $n\ge 3$, the $\Z_2$ code of Gray-mapped Kerdocks is 4-wise
independent.
\end{lemma}

\section{Fast List Decoding of Kerdock and Hankel Codes}\label{sec:recovery}

In this section we show how to perform quick list-decoding of the Hankel
code.  That is, we are given chosen-sampling access to a signal
$s$ and a parameter $k<N^c$ for some small $c\ge\Omega(1)$;
our goal is to find, with high probability, all Hankel codewords
$\varphi_{P,\ell}$ such that $|\dotprod{\varphi_{P,\ell}}{s}|^2 \ge
(1/k)\nerr{s}^2$, in time $\poly(k\log(N))$.

Our algorithm is a straightforward generalization of the algorithm
of~\cite{kushilevitz91learning}.  We do not give all the details of
this algorithm; instead, we refer the reader
to~\cite{kushilevitz91learning}.  Loosly speaking, the algorithm
in~\cite{kushilevitz91learning}
finds
$\ell$ for which $\varphi_\ell$ has large dot product with $s$ by
maintaining a set of candidates for the first $j$ bits of
$\ell$.  For $j=1,2,3,\ldots,n$, the algorithm extends each candidate
from $j-1$ to $j$ bits in all (two) ways, then tests each new
candidate.  The tests insure that the number of candidates remains
bounded, so the algorithm remains efficient.

Our algorithm will attempt to find first the $P$ matrix of each vector
$\varphi_{P,\ell}$ with $|\dotprod{\varphi_{P,\ell}}{s}|^2 \ge
(1/k)\nerr{s}^2$.  We will call such $P$ and such $\varphi_{P,\ell}$
{\em
heavy} for $s$.  Then the algorithm will find the $\ell$
part by demodulating out the contribution of $P$, and using the
algorithm in~\cite{kushilevitz91learning} to look for
heavy RM(1) vectors for $s\varphi_{P,0}^*$, where
$\varphi^*=1/(N\varphi)$ is the componentwise
complex conjugate of $\varphi$.  This strategy relies on
the fact that, up to normalization, $\dotprod{\varphi_{P,\ell}}{s}
=\dotprod{\varphi_{0,\ell}}{s\varphi_{P,0}^*}$, so $\varphi_{0,\ell}$
is heavy for $s\varphi_{P,0}^*$ when $\varphi_{P,\ell}$ is heavy for
$s$.

To find $P$, our algorithm follows the overall structure
of~\cite{kushilevitz91learning}.  For $j\le n$, we will maintain a
set of candidates for the upper-left $j$-by-$j$ submatrix of $P$.
The candidates will all be Hankel.  For each candidate, we will
consider extending it to a $(j+1)$-by-$(j+1)$ Hankel matrix, in one of
4 possible ways.  We then test each extended candidate in such a way
that, with high probability, all true candidates are kept (no false
negatives) but the total number of candidates kept is small enough
that our algorithm is efficient.  (We describe the retention criterion
and test in more detail below.)  Much of
the~\cite{kushilevitz91learning} algorithm works unchanged in our
context; we give few comments on those aspects and instead focus on
the changes necessary for the Hankel setting and the reasons our
algorithm works for Hankel but not for RM(2).  In particular, the
retention criterion and test we use are similar to that
in~\cite{kushilevitz91learning} and the guarantee of no false
negatives is similar; the main technical work is showing that there
are few (true or false) positives in the new context.  That is, the
analysis is
as follows:
\begin{enumerate}
\item Our algorithm is correct (finds all true candidates), by an
  analysis similar to~\cite{kushilevitz91learning}.
\item Our algorithm is efficient:
\begin{itemize}
\item As in~\cite{kushilevitz91learning}, the efficiency of
  our algorithm reduces to a non-algorithmic and
  non-probabilistic fact about the number of codewords with large dot
  product to the signal and the number of extensions of a $(j-1)$-by-$(j-1)$
  candidate to a $j$-by-$j$ candidate.
\item For the Hankel code in particular, we bound the number of
  codewords with large dot
  products and the number of extensions of a single candidate.  This
  is the only part of the proof where we will be formal since this is
  where our algorithm departs from previous work.
\end{itemize}
\end{enumerate}

Let us write $P\succeq \widetilde P$ if $\widetilde P$ is a square
submatrix
of $P$, consisting of the upper left $j$-by-$j$ corner of $P$ for some
$j$.
As
in~\cite{kushilevitz91learning}, we have an ideal testing criterion
for submatrices.
\begin{criterion}\label{crit:first}
A testing procedure keeps candidate $\widetilde P$ iff
there exists some $n$-by-$n$ matrix $P\succeq\widetilde P$ and some
$\ell\in \Z_2^n$ with
$\dotprodmag{\varphi_{P,\ell}}{s}^2 \ge (1/k)\nerr{s}^2$.
That is, the procedure keeps $\widetilde P$ iff there exists some unit-norm complex
number $c$, some $P\succeq\widetilde P$, and some $\ell$ with
$\Re\left(c\dotprod{\varphi_{P,\ell}}{s}\right) \ge (1/\sqrt{k})\nerr{s}$.
\end{criterion}

We will gradually rewrite and weaken this criterion in a sequence of
variations given below.  By ``weaken,'' we mean that a ``weaker'' criterion
will keep more matrices than a ``stronger'' criterion.   First,
for each $j$, for each string $y''$ of length $n-j$, and for
indeterminate $y'\in\Z_2^j$, define the restriction $(R_{y''}s)$ by
$(R_{y''}s)(y')=s(y'y'')$.  (Note that, if $\nerr{\varphi}=1$, then
$R_{y''}\varphi$ is {\em not} a unit vector.  We have
$\nerr{R_{y''}\varphi}^2=2^{j-n}$.)

Because $|\varphi_{P,\ell}|$ is constant, if
$\dotprod{R_{y''}\varphi_{P,\ell}}{R_{y''}s}$ is large,
then there must be many (small) contributions.  Formally:
\begin{lemma}
Suppose $|\varphi|\equiv 2^{-n/2}$, $\nerr{s}=1$, and
$\dotprodmag{\varphi}{s}\ge\sqrt{1/k}$.  Then, for each $j$, there are
at least $2^{n-j}/(4k)$ of $y''\in\Z_2^{n-j}$ such that
$\dotprodmag{R_{y''}\varphi}{R_{y''}s}\ge(1/\sqrt{4k})2^{j-n}$.
\end{lemma}
\begin{proof}
Suppose not.  Let $\psi$ be $\varphi$ restricted to the $y=y'y''$ with
$\dotprodmag{R_{y''}\varphi}{R_{y''}s}\ge(1/\sqrt{4k})2^{j-n}$, so
the support of $\psi$ has size less than $(2^n/(4k))$, and so
$\nerr{\psi}^2<1/(4k)$.
Then the at-most-$2^{n-j}$ possible $(y'')$'s with
$\dotprodmag{R_{y''}\varphi}{R_{y''}s}<(1/\sqrt{4k})2^{j-n}$
contribute a total of at most $1/\sqrt{4k}$ toward
$\dotprodmag{\varphi}{s}$, {\em i.e.},
$\dotprodmag{\varphi - \psi}{s}\le1/\sqrt{4k}$.  It follows that
\begin{eqnarray*}
\dotprodmag{\varphi}{s}
& \le & \dotprodmag{\psi}{s} + \dotprodmag{\varphi - \psi}{s}\\
& \le & \nerr{\psi}\nerr{s} + 1/\sqrt{4k}\\
&  <  & 1/\sqrt{4k} + 1/\sqrt{4k}\\
&  =  & 1/\sqrt{k},
\end{eqnarray*}
a contradiction.
\end{proof}

Thus we can weaken Criterion~\ref{crit:first} to:
\begin{criterion}\label{crit:most}
A testing procedure keeps candidate $\widetilde P$ iff
there exists some $n$-by-$n$ matrix $P\succeq\widetilde P$, a
unit-magnitude complex number $c$, and some
$\ell\in \Z_2^n$ such that, for at least $2^{n-j}/(4k)$ of
$y''\in\Z_2^{n-j}$ we have
$\Re\left(c\dotprod{R_{y''}\varphi_{P,\ell}}{R_{y''}s}\right)
\ge(1/(\sqrt{4k}))2^{j-n}\nerr{s}$.
\end{criterion}

Next, weaken Criterion~\ref{crit:most} to
\begin{criterion}\label{crit:each}
A testing procedure keeps candidate $\widetilde P$ iff
there exists some $n$-by-$n$ matrix $P\succeq\widetilde P$ and, for at
least $2^{n-j}/(4k)$ of $y''\in\Z_2^{n-j}$ there exists some unit-norm
$c_{y''}$ and some
$\ell_{y''}\in\Z_2^{n-j}$ with
$\Re\left(c_{y''}\dotprod{R_{y''}\varphi_{P,\ell_{y''}}}{R_{y''}s}\right)
\ge(1/(\sqrt{4k}))2^{j-n}\nerr{s}$.
\end{criterion}

Next, we will show that we need not search
over all possible $P\succeq\widetilde P$; a single fixed extension
$\widetilde P'$ will suffice.  (For example, $\widetilde P'$ might
extend $\widetilde P$ with zeros;  $\widetilde P'$ need not even be
Hankel.)   We will use the notation
$\widetilde P'$ in the sequel.
\begin{lemma}\label{lem:ext}
Fix parameter $n$, $j<n$, $j$-by-$j$ matrix $\widetilde P$, extension
$\widetilde P'\succeq \widetilde P$, and $y''\in\Z_2^{n-j}$.  Then
\[\{R_{y''}\varphi_{P,\ell}\,:\,
    P\succeq\widetilde P, \ell\in\Z_2^n\}
=
\{R_{y''}\varphi_{\widetilde P',\ell}\,:\,
  \ell\in\Z_2^n\}.\]
\end{lemma}

\begin{proof}
Write an extension $P$ to $\widetilde P$ as
\[
P=
\begin{pmatrix}
\widetilde P & P_1^T\\
P_1          & P_2
\end{pmatrix}
\]
and write $\ell^T=(\ell_1^T|\ell_2^T)$, where $\ell_1\in\Z_2^j$.
Then, at $y=y'y''$, we have
\begin{eqnarray*}
y^T P y + 2\ell^T y
& = & (y')^T \widetilde P y' + 2(y'')^T P_1 y' + (y'')^TP_2 y'' +
      2\ell_1^T y' + 2\ell_2^T y''\\
& = & (y')^T \widetilde P y' + 2((y'')^T P_1 + \ell_1^T) y' +
      ((y'')^TP_2 y'' + 2\ell_2^T y'').
\end{eqnarray*}
If we fix $y''$ but let $\ell$ vary, the expression $2((y'')^T P_1 +
\ell_1^T)$ varies over all of $2\Z_4^j$, whether or not we let $P_1$
vary.  Similarly, if we fix $y''$ but let the coefficient $c$ vary,
the expression $ci^{(y'')^TP_2 y'' +
2\ell_2^T y''}$ varies over unit-norm complex numbers, whether or not we let $P_2$ (and
$\ell_2$) vary.
\end{proof}

It follows that we can rewrite Criterion~\ref{crit:most} as
\begin{criterion}\label{crit:one-ext}
A testing procedure keeps candidate $\widetilde P$ iff for at
least $2^{n-j}/(4k)$ of $y''\in\Z_2^{n-j}$ there exists some unit-norm
$c_{y''}$ and some
$\ell_{y''}\in\Z_2^{n-j}$ with
\[\Re\left(c_{y''}\dotprod{R_{y''}\varphi_{\widetilde
    P',\ell_{y''}}}{R_{y''}s}\right)
\ge(1/(\sqrt{4k}))2^{j-n}\nerr{s}.\]
\end{criterion}

Finally, we will not be able to compute the test exactly, but we will
approximate with samples.  To that end, we need to have two
thresholds, with a gap.  Formally, we want the following criterion,
in which both the first and third cases represent a weakening,
compared with Criterion~\ref{crit:one-ext}:
\begin{criterion}\label{crit:HLDRC}
A testing procedure of a $j$-by-$j$ Hankel matrix $\widetilde P$ and signal $s$
with parameters $c_1$ and $c_2$ (determined below) behaves as follows.
\begin{itemize}
\item If for at least $2^{n-j}/(4k)$ of $y''\in\Z_2^{n-j}$ there
  exists some unit-norm $c_{y''}$ and some $\ell_{y''}\in\Z_2^{n-j}$
  with $\Re\left(c_{y''}\dotprod{R_{y''}\varphi_{\widetilde
  P',\ell_{y''}}} {R_{y''}s}\right)
\ge(1/(\sqrt{4k}))2^{j-n}\nerr{s}$, the procedure keeps $\widetilde P$
with high probability.
\item If only for less than
 $c_12^{n-j}/(4k)$ of $y''\in\Z_2^{n-j}$ does there exist some unit-norm
$c_{y''}$ and some
$\ell_{y''}\in\Z_2^{n-j}$ with
$\Re\left(c_{y''}\dotprod{R_{y''}\varphi_{\widetilde
    P',\ell_{y''}}}{R_{y''}s}\right)
\ge(1/(c_2\sqrt{4k}))2^{j-n}\nerr{s}$, the procedure drops $\widetilde P$
with high probability.
\item (The procedure may behave arbitrarily, otherwise.)
\end{itemize}
\end{criterion}

Our algorithm will also need an estimate for $\nerr{s}$.  Here we
simply assume that $\nerr{s}$ is known, say, up to the factor 2.
Alternatively, one might
assume the {\em dynamic range} of the problem is bounded, {\em i.e.},
that $1/M\le\nerr{s}\le M$ for some known $M$.  The algorithm could
then try
all $O(\log(M))$ possible $2^j$ in the range $1/M$ to $M$; one
of them is a factor-2 approximation to $\nerr{s}$.  This leads to an
extra factor of $\log(M)$ in some costs.  
One can also get an
appropriate approximation to $\nerr{s}$ from samples to $s$ without an
assumption about the dynamic range.  We omit
details; see~\cite{stoc-Fourier}.

We use the following straightforward efficient sampling algorithm to
implement Criterion~\ref{crit:HLDRC}, for which there exist suitable
$c_1$ and $c_2$:
\begin{algorithm}\label{algo:test}
Assuming $\nerr{s}$ is known to within a constant factor (that is
absorbed into $c_2$):
\begin{itemize}
\item For each $y''\in\Z_2^{n-j}$ such that $\nerr{R_{y''}s}^2 \le
  (40k/c_1)2^{j-n}\nerr{s}^2$, we can use the~\cite{kushilevitz91learning}
  algorithm to determine whether $\ell_{y''}$ and $c_{y''}$ for
  Criterion~\ref{crit:HLDRC} exist.
\item To determine whether at least $2^{n-j}/(4k)$ or at most
  $c_12^{n-j}/(4k)$ of the $y''\in\Z_2^{n-j}$ satisfy our condition,
  sample approximately $k/c_1$ of the
  $y''$'s; repeat
  to drive down failure probability.
  Note that
  there are at most $(c_1/10)2^{n-j}/(4k)$ possible $(y'')$'s where
  $\nerr{R_{y''}s}^2 > (40k/c_1)2^{j-n}\nerr{s}^2$.  The algorithm can
  behave arbitrarily on these $y''$ and still distinguish
  ``at most $(c_1)2^{n-j}/(4k)$'' from ``at least $2^{n-j}/(4k)$.''
\end{itemize}
\end{algorithm}

In summary, the following is a direct generalization of previous work
on RM(1) ({\em e.g.}, \cite{kushilevitz91learning}) concerning false
negatives, for which there is nothing special about RM(1) or Hankel:
\begin{proposition}
Fix parameter $n$ and signal $s$ of length $N=2^n$.
\begin{itemize}
\item For any $k$ and any $j\le n$, any procedure satisfying
  Criterion~\ref{crit:HLDRC} keeps, with high probability, all
  $j$-by-$j$ Hankel matrices
  $\widetilde P$ for which there is some $P\succeq \widetilde P$ and
  some $\ell\in\Z_2^n$ with
  $|\dotprod{s}{\varphi_\lambda}|^2\ge(1/k)\serr{s}^2$.
\item Algorithm~\ref{algo:test} satisfies
  Criterion~\ref{crit:HLDRC}.
\item Algorithm~\ref{algo:test} runs in time $\poly(k\log(N))$.
\end{itemize}
\end{proposition}

Thus we have shown that {\em each} call to Algorithm~\ref{algo:test},
to test a {\em single} candidate, is efficient.  We will call
Algorithm~\ref{algo:test} on many candidates as follows.
\begin{algorithm}\label{algo:ext}
Start with the exhaustive candidate set $C_1$ for $1$-by-$1$ matrices
$\widetilde P$.
For $j$ increasing from $1$ to $n-1$, extend all candidates in $C_j$
to $(j+1)$-by-$(j+1)$ Hankel matrices in all possible ways.  Call
Algorithm~\ref{algo:test} to test each candidate extension.
\end{algorithm}

It remains to show that the number of candidates $\widetilde P$ under
consideration remains under control.  Let $f(j)$ denote the number of
$j$-by-$j$ candidates considered.  Each candidate will be extended to a
$(j+1)$-by-$(j+1)$ Hankel matrix in all possible ways, getting $g(j+1)$
possible $(j+1)$-by-$(j+1)$ candidates.  Then
Criterion~\ref{crit:HLDRC} will be
applied to each candidate, reducing the number of candidates from
$g(j+1)$ to $f(j+1)$.  We need to bound both $f(j)$ and $g(j)$.  We
first bound $g(j+1)$ by $4f(j)$:

\begin{lemma}
Algorithm~\ref{algo:ext} constructs only four extensions to any
candidate.
\end{lemma}
\begin{proof}
Note that a $j$-by-$j$ candidate $\widetilde P$
extends to $(j+1)$-by-$(j+1)$ in only four ways, since there are only two new
possible bits, $a$ and $b$:
\[
\left(
\begin{array}{ccc|cc}
& & & & \\
 & \quad \widetilde P \quad & &  & \quad\quad\quad\\
& & & a & \\
\hline
 & & a & b & \quad\quad\quad\\
\vphantom{\Bigg|} & & &
\end{array}
\right).
\]
\end{proof}
Here we crucially use the fact that the candidates are Hankel.  If we
were to consider arbitrary $j$-by-$j$ RM(2) matrices, the number of
extensions would be $2^{j+1}$, which is prohibitive.

Thus it suffices to bound $f(j)$ by a polynomial in $k$, uniformly
for all $j$.  So we need to bound the number of 
$\widetilde P$'s that are $(c_2/k)$-heavy on more than
$c_1(k)2^{n-j}/k$ of the $y''\in\Z_2^{n-j}$, where we call a candidate
$\widetilde P$ {\em
  $h$-heavy on $y''$} if $\widetilde P$ extends to some $P$ with some
$\ell_{y''}$ satisfying
$\dotprodmag{R_{y''}\varphi_{P,\ell_{y''}}}{R_{y''}s} \ge
\sqrt{h}2^{j-n}\nerr{s}$.  The other candidates are dropped by our criterion.

\begin{figure}
\begin{center}
\begin{picture}(220,100)(0,10)
\put(20,100){\makebox(0,0)[l]{$y''$}}
\put(40,100){\vector(1,0){180}}
\put(220,100){\makebox(0,0)[l]{\quad$\le 2^{n-j}$}}

\put(0,80){\makebox(0,0)[b]{$\widetilde P$}}
\put(0,75){\vector(0,-1){50}}

\put(20,90){\line(1,0){200}}
\put(20,10){\line(1,0){200}}

\put(20,90){\line(0,-1){80}}
\put(220,90){\line(0,-1){80}}

\put(100,20){\makebox(0,0)[c]{$(c_2/k)$-heavy}}

\put(35,80){\makebox(0,0)[c]{$\checkmark$}}
\put(45,80){\makebox(0,0)[c]{$\checkmark$}}
\put(65,80){\makebox(0,0)[c]{$\checkmark$}}
\put(80,80){\makebox(0,0)[c]{$\checkmark$}}
\put(90,80){\makebox(0,0)[c]{$\checkmark$}}
\put(120,80){\makebox(0,0)[c]{$\checkmark$}}
\put(135,80){\makebox(0,0)[c]{$\checkmark$}}
\put(150,80){\makebox(0,0)[c]{$\checkmark$}}
\put(170,80){\makebox(0,0)[c]{$\checkmark$}}

\put(40,60){\makebox(0,0)[c]{$\checkmark$}}
\put(110,50){\makebox(0,0)[c]{$\checkmark$}}
\put(190,70){\makebox(0,0)[c]{$\checkmark$}}
\end{picture}
\end{center}
\caption{Bounding the number of $\widetilde P$'s that are heavy on
  many $(y'')$'s.  Label columns by $(y'')$'s for which
  $\nerr{R_{y''}s}^2 \le (k/c_3)(2^{j-n})\nerr{s}^2$ and label rows by
  $\widetilde P$'s; put a checkmark at $(y'',\widetilde P)$ if
  $\widetilde P$ is $(c_2/k)$-heavy for $y''$.  We will show below that
  there are few checkmarks in any column; it follows that there are
  few rows with checkmarks in many columns.}\label{fig:minmax}
\end{figure}
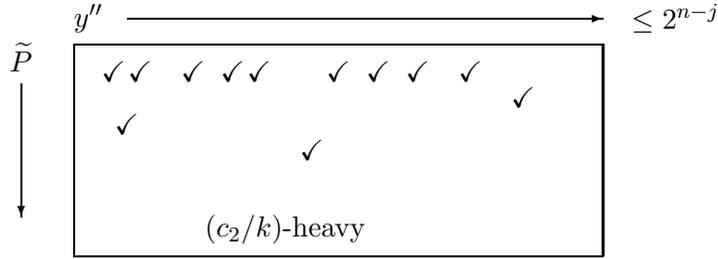

As in previous work, it suffices to bound the number of candidates
$\widetilde P$ for {\em each} $y''$ and then do an averaging
argument.  There are
at most $(c_3/k)2^{n-j}$ possible $(y'')$'s for which $\nerr{R_{y''}s}^2
\ge (k/c_3)2^{j-n}\nerr{s}^2$ and, for constant
$c_3$ related to $c_1$, these
$(y'')$'s can be ignored in determining whether $\widetilde P$
satisfies the condition of~Criterion~\ref{crit:HLDRC} on at least
$(1/(4k))2^{n-j}$ or
at most $(c_1/(4k))2^{n-j}$ of the $(y'')$'s.  So, henceforth, consider
only $y''$ for which $\nerr{R_{y''}s}^2 < (k/c_3)2^{j-n}\nerr{s}^2$.
Below, for all such $y''$, we will bound, by $B_k\le\poly(k)$, the
number of $\widetilde P$ that are $(c_2/k)$-heavy on $y''$.
Summing over 
at most $2^{n-j}$ possible $(y'')$'s, there are at most $B_k2^{n-j}$
pairs $(\widetilde P,y'')$ where $\widetilde P$ is
$(c_2/k)$-heavy for $y''$.  Thus there can be at most
$B_k\cdot(k/c_1)\le\poly(k)$ possible $\widetilde P$'s that are
$(c_2/k)$-heavy
on at least $(c_1/k)2^{n-j}$ of the $(y'')$'s; {\em i.e.}, at any
stage $j$, there are at most $f(j)\le\poly(k)$ possible candidates
considered by our algorithm.  See Figure~\ref{fig:minmax}.

Thus we have, from previous work and without specific consideration of
the Hankel code,
\begin{proposition}
Fix signal $s$ of length $N=2^n$, fix parameter $k$, and fix $j\le n$.
Suppose, for each $y''\in\Z_2^{n-j}$ with $\nerr{R_{y''}s}^2 \le
(k/c_3)\nerr{R_{y''}s}^2$, there are at most $\poly(k)$
possible $j$-by-$j$ Hankel matrices $\widetilde P$ that are
$(c_2/k)$-heavy for $y''$.  Then there are at most $\poly(k)$
possible $\widetilde P$ that
are kept by our algorithm.
\end{proposition}

Finally, we now proceed to Hankel-specific analysis.  To simplify notation, and
without loss of generality, we drop all previous constants.  It
suffices to show that there are at most $\poly(k)$ Hankel matrices $P$
such that there exists an $\ell$ with
$|\dotprod{s}{\varphi_{P,\ell}}|^2\ge(1/k)\nerr{s}^2$.

In an orthonormal basis, by the Parseval Equality, there can be at
most $k$ vectors $\varphi$ with
$|\dotprod{s}{\varphi}|^2\ge(1/k)\nerr{s}^2$.  Similarly, in a
$\mu$-incoherent dictionary, {\em i.e.}, a set of vectors with all dot
product magnitudes bounded above by $\mu$, if $\mu k$ is at most some
constant $c_4\approx 1/6$, then there are at most
$O(k)$ such
$\lambda$'s~\cite{TGMS03:Improved-Sparse,gilbert03approximation}.
Hankel,
however, is not a $\mu$-incoherent set for small $\mu$, because there
are pairs $P$ and $P'$ of Hankel matrices that differ by a low-rank
matrix, whence the corresponding vectors $\varphi_{P,0}$ and
$\varphi_{P',0}$ have large dot product.  Nevertheless, we show that,
for each $P$, the number of $P'$ such that $P+P'$ has low rank is
small.  We then show that the set of Hankel codewords works like an
orthonormal basis or an incoherent set, in the sense that there may be
at most $\poly(k)$ Hankel $\lambda$'s with
$|\dotprod{s}{\varphi_\lambda}|^2\ge(1/k)\nerr{s}^2$.

We now proceed formally.  This proceeds in a sequence of lemmas along
with Dickson's Theorem (Theorem~\ref{thm:dickson}), all
of which have proofs that are elementary or found in existing work.

\begin{lemma}\label{lem:incoh}
There is a constant $c_4$ such that, for any incoherence parameter
$\mu$, $0\le\mu\le 1$ any $k$, and any signal $s$, if $\mu k\le c_4$,
there are at most $O(k)$
vectors in any set $A$ such that both of the following hold:
\begin{itemize}
\item For all $\varphi\ne\varphi'\in A$, we have
  $|\dotprod{\varphi}{\varphi'}|\le\mu$.
\item For all $\varphi\in A$, we have
  $|\dotprod{s}{\varphi}|^2\ge(1/k)\nerr{s}^2$.
\end{itemize}
\end{lemma}
\begin{proof}
This essentially follows
from~\cite{TGMS03:Improved-Sparse,gilbert03approximation}; we include
a sketch of the proof with possibly different constants.  Suppose,
toward a contradiction, there are $\ell>4k$ vectors in $A$; wlog,
$\ell=4k+1$, since we can discard the remaining vectors.  We may
assume that $s=\sum_j a_j\varphi_j$ lies in the span
of $A=\{\varphi_j\}$.  The idea is to show that
$\nerr{s}^2\approx\sum_j|a_j|^2$ and
$\dotprodmag{s}{\varphi_j}^2\approx |a_j|^2$, so
that an approximate Parseval equality holds.  First,
\begin{eqnarray*}
\nerr{s}^2
&  =  & \dotprod{\sum_j a_j\varphi}{\sum_{j'} a_{j'}\varphi}\\
& \ge & \sum_j |a_j|^2 - \mu\left|\sum_{j\ne j'} a_j\overline{a_{j'}}\right|\\
& \ge  & \sum_j |a_j|^2 - \mu\left|\sum_{j} a_j\right|^2\\
& \ge  & \sum_j |a_j|^2 - \mu(4k+1)\sum_{j} |a_j|^2,
\end{eqnarray*}
by Cauchy-Schwarz, so that, for some $c$,
\begin{equation}
\sum_{j} |a_j|^2\le(1 + c\mu k)\nerr{s}^2.\label{eq:approx_parseval}
\end{equation}
On the other hand, for each $j$,
\begin{eqnarray*}
\dotprodmag{s}{\varphi_j}
&  =  & \left|a_j+\sum_{j'\ne j}
          a_{j'}\dotprod{\varphi_{j'}}{\varphi_j}\right|\\
& \le & |a_j|+\mu\sum_{j'\ne j}|a_{j'}|\\
& \le & |a_j|+\mu\sqrt{4k\sum_{j'\ne j}|a_{j'}|^2}\\
&  =  & |a_j|+O(\mu k)(1/\sqrt{k})\nerr{s},
\end{eqnarray*}
so that, for some $c'$ we hav $|a_j|\ge\dotprodmag{s}{\varphi_j}-O(\mu
k)(1/\sqrt{k})\nerr{s}$, and so
\[|a_j|^2\ge(1-c'\mu k)^2(1/k)\nerr{s}^2.\]
Summing over all $\ell=4k+1$ terms, we get
\[\sum_j |a_j|^2 \ge (1- c'\mu k)^2(\ell/k)\nerr{s}^2,\]
so that, with~(\ref{eq:approx_parseval}), we get
$(1- c'\mu k)^2(\ell/k)\le(1 + c\mu k)$, or
$\ell\le k(1 + c\mu k)(1 - c'\mu k)^{-2}$.  Thus, if $k\mu$ is
a sufficiently small constant, we get $\ell\le 4k$, a contradiction.
\end{proof}

\begin{lemma}\label{lem:few}
For any constant $c_4$, there are just $L_k\le \poly(k)$ Hankel
matrices of rank at most $2\log(k/c_4)$.  Equivalently, for each Hankel
$P$, there are at most $L_k$ Hankels $P'$ with
$\rank(P+P')\le2\log(k/c_4)$.
\end{lemma}
\begin{proof}
Suppose Hankel matrix $P$ has rank $r$.  We claim that $O(r)$ binary
parameters determine the top half of the matrix (above the main
reverse diagonal).  Another $O(r)$ parameters determine the bottom
half, whence the number of such matrices is $2^{O(r)}$.  The result
follows.

Write the $(r+1)$'st column as a linear combination $C$ of the first
$r$ columns.  We claim that $C$ and the first $r$ entries
$p_0,p_1,\ldots,p_{r-1}$ in the top row determine the top half of the
matrix.  Determine $p_r$ from $p_0,p_1,\ldots,p_{r-1}$ and $C$ applied
to the top row (row 0).  Then, having determined $p_r$, determine
$p_{r+1}$ from $C$ applied to the first $r$ entries in row 1, {\em
  i.e.}, $p_1,p_2,\ldots,p_r$.  Proceed to determine $p_{r+2}$ from
$C$ applied to the first $r$ entries in row 2, {\em
  i.e.}, $p_2,p_3\ldots,p_{r+1}$.  The general statement follows by
induction.

For example, suppose Hankel $P$ has rank three, the first three
reverse diagonal parameters are $a,b,c$, and column 3 is the linear
combination $C$ of columns $0,1,2$.  Then, in
\[
P=\left(
\begin{array}{ccc|c|c}
a & b & c & d & e\\
b & c & d & e & f\\
c & d & e & f & g\\
d & e & f & g & h\\
\vdots\\
\end{array}
\right),
\]
we get $d$ in row 0, column 3 from $a,b,c$ by applying $C$ in row 0.
Now knowing $d$ in addition to $a,b,c$, we get $e$ in row 1, column 3
by applying $C$ to $b,c,d$ in row 1.  We get $f$ in row 2, column 3 by
applying $C$ to $c,d,e$, etc.
\end{proof}

In intermediate stages of our algorithm, we need to bound only the
number of Hankel {\em matrices} $P$ that are considered.  In the output,
however, 
we need to bound the total number of Hankel {\em codewords} output, {\em
  i.e.}, the number of pairs $(P,\ell)$.  We give the latter stronger
statement
in this summary theorem.

\begin{theorem}\label{thm:few_hankel}
For any signal $s$, there are at most $\poly(k)$ Hankel codewords
$\varphi_{P,\ell}$ with
\[|\dotprod{s}{\varphi_{P,\ell}}|^2\ge(1/k)\nerr{s}^2.\]
\end{theorem}
\begin{proof}
Suppose there are at least $q$ Hankel codewords $\varphi_{P,\ell}$
with
$|\dotprod{s}{\varphi_{P,\ell}}|^2\ge(1/k)\nerr{s}^2$.  For fixed $P$,
the set $\{\varphi_{P,\ell}:\ell\}$ is an orthonormal basis, so there
are at most $k$ possible $\ell$'s for each $P$ with
$|\dotprod{s}{\varphi_{P,\ell}}|^2\ge(1/k)\nerr{s}^2$.  Thus there are
at least $q/k$ matrices $P$ with at least one $\ell$ satisfying
$|\dotprod{s}{\varphi_{P,\ell}}|^2\ge(1/k)\nerr{s}^2$.  By
Lemma~\ref{lem:few}, there is a set $Q$ of size $|Q|\ge q/(kL_k)$
matrices $P$ having an $\ell$ satisfying
$|\dotprod{s}{\varphi_{P,\ell}}|^2\ge(1/k)\nerr{s}^2$ and with
$\rank(P+P')\ge2\log(k/c_4)$ for all $P\ne P'\in Q$.  By
Theorem~\ref{thm:dickson}, for any $P\ne P'\in Q$ and their
corresponding $\ell$ and $\ell'$, we have
$|\dotprod{\varphi_{P,\ell}}{\varphi_{P',\ell'}}|\le (c_4/k)$.  By
Lemma~\ref{lem:incoh}, $|Q|\le O(k)$.  It follows that
$q\le\poly(k)$.
\end{proof}

In summary, we have our main theorem.
\begin{theorem}
Let $\{\varphi_\lambda\}$ denote the Hankel code.  There is an
algorithm that, given parameter $k$ and
chosen-sampling access to a signal $s\in\C^N$, finds, in time
$\poly(k\log(N))$, a list
containing all $\lambda$ with
$|\dotprod{s}{\varphi_\lambda}|^2\ge(1/k)\nerr{s}^2$.
\end{theorem}

\section{Conclusion}\label{sec:conc}

\subsection{Corollaries}

A list-decoding algorithm for the Hankel code immediately gives a
list-decoding algorithm for the Kerdock subcode.  Since the Kerdock
code is $(1/\sqrt{N})$-incoherent, we immediately get a sparse
recovery algorithm for Kerdock,
using~\cite{TGMS03:Improved-Sparse,gilbert03approximation}.  That is:
\begin{corollary}
Let $\{\varphi_\lambda\}$ denote a Kerdock code that is a subset of a
Hankel code.  There is an algorithm that, given parameter $k$ and
chosen-sampling access to a signal $s\in\C^N$, finds, in time
$\poly(k\log(N))$, a list
containing all $\lambda$ with
$|\dotprod{s}{\varphi_\lambda}|^2\ge(1/k)\nerr{s}^2$.
\end{corollary}

\begin{corollary}
Let $\{\varphi_\lambda\}$ denote a Kerdock code that is a subset of a
Hankel code.  There is an algorithm that, given parameters
$k<1/(6\sqrt{N})$ and $\epsilon>0$ and chosen-sampling access
to a signal $s\in\C^N$, finds, in time $\poly(k\log(N)/\epsilon)$, a
set $\Lambda$ of size $k$ and coefficients $c_\lambda$ ({\em i.e.}, a
$k$-term approximation $\widetilde s=\sum_{\lambda\in\Lambda}
c_\lambda\varphi_\lambda$)
with $\nerr{\widetilde s - s}^2
\le(1+\epsilon+k^2/\sqrt{N})\nerr{s_k-s}^2$, where $s_k$ is the best
$k$-term Kerdock approximation to $s$.
\end{corollary}

\subsection{Improvements}

The cost of our Hankel recovery algorithm is polynomial in $k$, but
high.  In Lemma~\ref{lem:few}, we show only that there are at most
$2^{4r}$ Hankel matrices of rank $r$, whence, for each Hankel $P$,
there are at most $2^{4r}=k^8$ Hankel matrices $P'\ne P$ with
$|\dotprod{\varphi_{P,\ell}}{\varphi_{P',\ell'}}|>(1/k)=2^{-r/2}$.
This means we bound
the time cost of our algorithm at $k^c$ for $c$ an integer somewhat
larger than 8.  We make a few comments:
\begin{itemize}
\item It is easy to see that there are at least $\Omega(k^4)$ Hankel
  matrices of rank $1/k$.  If we really want to
  list-decode
  Hankel rather than Kerdock, the size of the output can really be at
  least approximately $k^5$.  Our runtime of $k^c$ will be
  approximately quadratic in the size of the output, which may be
  acceptable in some contexts.\footnote{The
  $\approx k^5$ output Hankel codewords come in (possibly overlapping)
  clusters of approximately $k^4$ vectors each, so there are at most
  approximately $k$ clusters.  One might hope to produce a compressed
  representation of the output in less time than it takes to write out
  the output uncompressed.  Note, however, that the boundaries of the
  clusters are generally not smooth, so it will not suffice to output
  the cluster centers.}
\item A tighter analysis of the way the top and bottom halves of the
  matrix fit together may bound the number of rank-$(1/k)$ Hankels more
  tightly than $k^8$.
\item We have begun to investigate an alternative algorithm that
  exploits the fact that the restriction of a Kerdock codeword to a
  sub{\em field} is a smaller instance of a Kerdock codeword.  This
  algorithm is much faster as a list-decoding algorithm for Kerdock
  only, since it doesn't keep so many candidates.  But the paradigm of
  bit-by-bit extensions in the algorithm
  of~\cite{kushilevitz91learning} and
  Algorithm~\ref{algo:ext} does not work for subfields.
\end{itemize}
Faster algorithms to list-decode Kerdock codes will be the subject of
future work.

Other future work will include extensions to the Delsarte-Goethals
hierarchy of codes between RM(1) and RM(2).  As one ascends the
hierarchy, the size of the code increases as the the maximum dot
product increases.

\section*{Acknowledgment}
We thank Joel Lepak, Muthu Muthukrishnan and Alex Samorodnitsky for
helpful
discussions.

\bibliographystyle{alpha}
\bibliography{kerdockbib}

\newcommand{\etalchar}[1]{$^{#1}$}
\begin{thebibliography}{AKK{\etalchar{+}}03}

\bibitem[AGS03]{akavia:goldwasser}
A.~Akavia, S.~Goldwasser, and S.~Safra.
\newblock Proving hard-core predicates using list decoding.
\newblock In {\em in Proc. of the 44th Annual IEEE Symposium on Foundations of
  Computer Science}, 2003.

\bibitem[AKK{\etalchar{+}}03]{alon:kaufman}
Noga Alon, Tali Kaufman, Michael Krivelevich, Simon Litsyn, and Dana Ron.
\newblock Testing low-degree polynomials over $gf(2)$.
\newblock In {\em Proc. of APPROX+RANDOM}, 2003.

\bibitem[AS97]{arora97improved}
Sanjeev Arora and Madhu Sudan.
\newblock Improved low-degree testing and its applications.
\newblock In {\em Proc.~29th ACM symposium on Theory of computing}, pages
  485--495, 1997.

\bibitem[CCKS97]{calderbank:cameron:kantor:seidel}
A.R. Calderbank, P.J. Cameron, W.M. Kantor, and J.J. Seidel.
\newblock $\mathbb{Z}_4$-kerdock codes, orthogonal spreads, and extremal
  euclidean line-sets.
\newblock {\em Proc. London Math. Soc. (75)}, pages 436--480, 1997.

\bibitem[CGL{\etalchar{+}}05]{CGLMS}
A.~R. Calderbank, A.~Gilbert, K.~Levchenko, S.~Muthukrishnan, and M.~Strauss.
\newblock Improved range-summable random variable construction algorithms.
\newblock In {\em SODA '05: Proceedings of the sixteenth annual ACM-SIAM
  symposium on Discrete algorithms}, pages 840--849, Philadelphia, PA, USA,
  2005. Society for Industrial and Applied Mathematics.

\bibitem[GGI{\etalchar{+}}02]{stoc-Fourier}
Anna~C. Gilbert, Sudipto Guha, Piotr Indyk, S.~Muthukrishnan, and Martin
  Strauss.
\newblock Near-optimal sparse {Fourier} representations via sampling.
\newblock In {\em Proc. 34th ACM Symposium on Theory of Computing}, pages
  152--161, 2002.

\bibitem[GL89]{GL}
O.~Goldreich and L.~Levin.
\newblock A hard-core predicate for all one-way functions.
\newblock In {\em Proc. 21st ACM Symposium on Theory of Computing}, pages
  25--32. ACM, 1989.

\bibitem[GMS03]{gilbert03approximation}
A.~Gilbert, S.~Muthukrishnan, and M.~Strauss.
\newblock Approximation of functions over redundant dictionaries using
  coherence, 2003.

\bibitem[HCM06]{howard:calderbank:moran}
S.~D. Howard, A.~R. Calderbank, and W.~Moran.
\newblock The finite heisenberg-weyl groups in radar and communications.
\newblock {\em EURASIP Journal on Applied Signal Processing}, 2006:Article ID
  85685, 12 pages, 2006.

\bibitem[HKC{\etalchar{+}}94]{hammons:calderbank}
A.~R. Hammons, P.V. Kumar, A.~R. Calderbank, N.J.A. Sloane, and P.~Sol\'e.
\newblock The $\mathbb{Z}_4$-linearity of {Kerdock}, {Preparata}, {Goethals},
  and related codes.
\newblock {\em IEEE Trans. on Information Theory}, 40(2):301--319, 1994.

\bibitem[HSP06]{HSP}
R.~W. Heath, T.~Strohmer, and A.~J. Paulraj.
\newblock On quasi-orthogonal signatures for {CDMA} systems.
\newblock {\em {IEEE} Transactions on Information Theory}, 52(3), March 2006.

\bibitem[Ind00]{Ind00:High-Dimensional-Computational}
P.~Indyk.
\newblock {\em High-Dimensional Computational Geometry}.
\newblock PhD thesis, Stanford, 2000.

\bibitem[KL05]{Kaufman05}
T.~Kaufman and S.~Litsyn.
\newblock Almost orthogonal linear codes are locally.
\newblock In {\em Proc.~46th Foundations of Computer Science}, pages 317--326.
  IEEE, 2005.

\bibitem[KM91]{kushilevitz91learning}
Eyal Kushilevitz and Yishay Mansour.
\newblock Learning decision trees using the {Fourier} spectrum.
\newblock pages 455--464, 1991.

\bibitem[Man95]{Mansour}
Y.~Mansour.
\newblock Randomized interpolation and approximationof sparse polynomials.
\newblock {\em SIAM Journal on Computing}, 24(2):357--368, 1995.

\bibitem[MS77]{macwilliams:sloane}
F.J. MacWilliams and N.J.A. Sloane.
\newblock {\em The Theory of Error-Correcting Codes}.
\newblock North-Holland, 1977.

\bibitem[Sud]{madhu-hp}
Madhu Sudan.
\newblock Algorithmic introduction to coding theory.
\newblock Course Home Page.
\newblock \url{http://theory.lcs.mit.edu/~madhu/FT01/}.

\bibitem[Sud00]{sudan:list}
Madhu Sudan.
\newblock List decoding: Algorithms and applications.
\newblock In {\em Proc. of the International Conference IFIP TCS 2000}, 2000.

\bibitem[Sud01]{sudan:survey}
Madhu Sudan.
\newblock Coding theory: Tutorial and survey.
\newblock In {\em Proc. of the 42nd Annual Symposium on Foundations of Computer
  Science}, 2001.

\bibitem[TGMS03]{TGMS03:Improved-Sparse}
J.~A. Tropp, A.~C. Gilbert, S.~Muthukrishnan, and M.~J. Strauss.
\newblock Improved sparse approximation over quasi-incoherent dictionaries.
\newblock In {\em Proc. of the 2003 IEEE International Conference on Image
  Processing}, Barcelona, 2003.

\end{thebibliography}

\end{document}